\documentclass[10pt,english]{article}
\usepackage[T1]{fontenc}
\usepackage[latin1]{inputenc}
\usepackage{geometry}
\usepackage{graphicx}

\makeatletter

\providecommand{\tabularnewline}{\\}

\usepackage{babel}
\makeatother
\begin{document}

\title{Comparative studies of the magnetic dipole and electric quadrupole
hyperfine constants for the ground and low lying excited states of
$^{25}Mg^{+}$ }

\author{Chiranjib Sur, Bijaya K. Sahoo, Rajat K Chaudhuri, B. P. Das \\
\emph{Non-Accelerator Particle Physics Group, Indian Institute of
Astrophysics, }\\
\emph{Bangalore - 560 034, India}\\
 D. Mukherjee\\
\emph{Indian Association for the Cultivation of Science, Kolkata -
700 032, India}}

\date{(Received date, Accepted date)}

\maketitle
\begin{abstract}
We have employed the relativistic coupled cluster theory to calculate
the magnetic dipole and electric quadrupole hyperfine constants for
the ground and low lying excited states of singly ionized magnesium.
Comparison with experimental and the other theoretical results are
done and predictions are also made for a few low lying excited states
which could be of interest. We have made comparative studies of the
important many body effects contributing to the hyperfine constants
for the different states of the ion.

\textbf{PACS number(s).} : 31.15.Ar, 31.15.Dv, 31.25.Jf, 32.10.Fn
\end{abstract}

\section{\label{intro}Introduction}

The hyperfine interactions in alkali metal atoms and alkaline earth
ions have been of interest for quite a long time \cite{arimondo,mg+exp}.
A number of theoretical investigations including some based on relativistic
many-body theory have been performed \cite{tpdas-mg+,johnson-hyp,fischer}
and they compare reasonably well with experiments. Some of the theoretically
predicted values could be of experimental interest with the advent
of high precision techniques \cite{mod-phys-book} involving trapped
and laser cooled atoms \cite{miskatul} and ions \cite{mg+exp}. 

The high precision calculations of different properties of many-electron
atoms requires accurate wave-functions in the nuclear region as well
as the region far from the nucleus. The study of properties like hyperfine
constants requires the former. Since hyperfine interactions are sensitive
to electron correlations, the determination of atomic hyperfine constant
provides an important test for \emph{ab initio} atomic structure theory
\cite{lindgren-book}.

In this paper we have carried out \emph{ab-initio} calculations of
the magnetic dipole and electric quadrupole hyperfine constants and
compared the many-body effects for the ground as well as some excited
states. Section \ref{theory} provides the necessary theoretical background
to the magnetic dipole ($A$) and the electric quadrupole ($B$) hyperfine
constants. Section \ref{cc-overview} gives an overview of the coupled-cluster
theory and its application to this specific problem. Finally in section
\ref{results} the results of our calculations are presented and discussed.

\section{\label{theory}Theory}

The interaction between the various moments of the nucleus and the
electrons of an atom are collectively referred to as hyperfine interactions
\cite{lindgren-book}. In this paper we shall consider the interactions
between the atomic electrons with the nuclear magnetic dipole and
electric quadrupole moments. Nuclear spin gives rise to a nuclear
magnetic dipole moment and the departure from spherical charge distribution
in the nucleus produces an electric quadrupole moment. 

The hyperfine interaction is given by \cite{cheng}

\begin{equation}
H_{hfs}=\sum_{k}M^{(k)}\cdot T^{(k)},\label{eqn-1}\end{equation}
where $M^{(k)}$ and $T^{(k)}$ are spherical tensors of rank $k$,
which corresponds to nuclear and electronic parts of the interaction
respectively. The lowest $k=0$ order represents the interaction of
the electron with the spherical part of the nuclear charge distribution.
The eigenstates of the atomic Hamiltonian in the presence of a hyperfine
interaction are denoted by $\left|IJFM_{F}\right\rangle $. Here $\mathbf{I}$
and $\mathbf{J}$ are the total angular angular momentum for the nucleus
and the electron state, respectively, and $\mathbf{F}=\mathbf{I}+\mathbf{J}$
with the projection $M_{F}$. 

In the first order perturbation theory, the energy corresponding to
the hyperfine interaction of the fine structure state $\left|JM_{J}\right\rangle $
are the expectation values of $H_{hfs}$ such that

\begin{equation}
\begin{array}{ccc}
W(J) & = & \left\langle IJFM_{F}\right|{\displaystyle \sum_{k}}M^{(k)}\cdot T^{(k)}\left|IJFM_{F}\right\rangle \\
 & = & {\displaystyle \sum_{k}}(-1)^{I+J+F}\left\{ \begin{array}{ccc}
I & J & F\\
J & I & k\end{array}\right\} \left\langle I\right\Vert M^{(k)}\left\Vert I\right\rangle \left\langle J\right\Vert T^{(k)}\left\Vert J\right\rangle \end{array}\label{eqn-2}\end{equation}

The $k=1$ term describes the magnetic dipole coupling of the nuclear
magnetic moment with the magnetic field created by the electron at
the position of the nucleus. The nuclear dipole moment $\mu_{I}$
is defined (in units of Bohr magneton $\mu_{N}$) as 

\begin{equation}
\mu_{I}\mu_{N}=\left\langle II\right|M_{0}^{(1)}\left|II\right\rangle =\left(\begin{array}{ccc}
I & 1 & I\\
-I & 0 & I\end{array}\right)\left\langle I\right\Vert M^{(1)}\left\Vert I\right\rangle \label{eqn-3}\end{equation}
and the operator $T_{q}^{(1)}$is given by\cite{lind-case-st}

\begin{equation}
T_{q}^{(1)}=\sum_{q}t_{q}^{(1)}=\sum_{j}-ie\sqrt{\frac{8\pi}{3}}r_{j}^{-2}\overrightarrow{\alpha_{j}}\cdot\mathbf{Y}_{1q}^{(0)}(\widehat{r_{j}}).\label{eqn-4}\end{equation}
Here $\overrightarrow{\alpha}$ is the Dirac matrix and $\mathbf{Y}_{kq}^{\lambda}$
is the vector spherical harmonics. In Eq.(\ref{eqn-4}) the index
$j$ refers to the $j$-th electron of the atom and $e$ is the magnitude
of the electronic charge. The magnetic dipole hyperfine constant $A$
is defined as 

\begin{equation}
A=\mu_{N}\left(\frac{\mu_{I}}{I}\right)\frac{\left\langle J\right\Vert T^{(1)}\left\Vert J\right\rangle }{\sqrt{J(J+1)(2J+1)}},\label{eqn-5}\end{equation}
and the corresponding magnetic dipole hyperfine energy $W_{M1}$ is
given by

\begin{equation}
W_{M1}=A\left\langle I\cdot J\right\rangle =A\frac{K}{2},\label{eqn-6}\end{equation}
where $K=F(F+1)-I(I+1)-J(J+1)$.

The second order term in the hyperfine interaction is the electric
quadrupole part. The electric quadrupole hyperfine constant is defined
by putting $k=2$ in Eq. (\ref{eqn-2}). The nuclear quadrupole moment
is defined as

\begin{equation}
T_{q}^{(2)}=\sum_{q}t_{q}^{(2)}=\sum_{j}-er_{j}^{-3}C_{q}^{(2)}(\widehat{r_{j}}),\label{eqn-7}\end{equation}
Here, $C_{q}^{(k)}=\sqrt{\frac{4\pi}{(2k+1)}}Y_{kq}$, with $Y_{kq}$
being the spherical harmonic. Hence the electric quadrupole hyperfine
constant $B$ is

\begin{equation}
B=2eQ\left[\frac{2J(2J-1)}{(2J+1)(2J+2)(2J+3)}\right]^{1/2}\left\langle J\right\Vert T^{(2)}\left\Vert J\right\rangle ,\label{eqn-8}\end{equation}
and the corresponding electric quadrupole hyperfine energy $W_{E2}$
is given by

\begin{equation}
W_{E2}=\frac{B}{2}\frac{3K(K+1)-4I(I+1)J(J+1)}{2I(2I-1)2J(2J-1)}.\label{eqn-9}\end{equation}
In Eq.( \ref{eqn-4} and \ref{eqn-7} ) $t_{q}^{(k)}$are the single
particle reduced matrix element for the electronic part. The reductions
of the single particle matrix elements into angular factors and radial
integral are straightforward by means of using the Wigner Eckart theorem.
These single particle reduced matrix elements are given by

\begin{equation}
\left\langle \kappa\right\Vert t_{q}^{(1)}\left\Vert \kappa^{\prime}\right\rangle =-\left\langle \kappa\right\Vert C_{q}^{(1)}\left\Vert \kappa^{\prime}\right\rangle (\kappa+\kappa^{\prime})\int dr\frac{\left(P_{\kappa}Q_{\kappa^{\prime}}+Q_{\kappa}P_{\kappa^{\prime}}\right)}{r^{2}}\label{eqn-10}\end{equation}
and

\begin{equation}
\left\langle \kappa\right\Vert t_{q}^{(2)}\left\Vert \kappa^{\prime}\right\rangle =-\left\langle \kappa\right\Vert C_{q}^{(2)}\left\Vert \kappa^{\prime}\right\rangle \int dr\frac{\left(P_{\kappa}P_{\kappa^{\prime}}+Q_{\kappa}Q_{\kappa^{\prime}}\right)}{r^{3}},\label{eqn-11}\end{equation}
where $\left\langle \kappa\right\Vert C_{q}^{(k)}\left\Vert \kappa^{\prime}\right\rangle $
is the reduced matrix element of the spherical tensor and is equal
to

\[
(-1)^{j+1/2}\sqrt{(2j+1)(2j^{\prime}+1)}\left(\begin{array}{ccc}
j & k & j^{\prime}\\
\frac{1}{2} & 0 & -\frac{1}{2}\end{array}\right)\pi(l,k,l^{\prime}),\]
with

\[
\pi(l,k,l^{\prime})=\left\{ \begin{array}{c}
\begin{array}{cc}
1 & \mathrm{if}\: l+k+l^{\prime}\,\,\mathrm{even}\\
0 & \mathrm{otherwise}\end{array}\end{array}\right.\]

Here the single particle orbitals are expressed in terms of the Dirac
spinors with $P_{i}$ and $Q_{i}$ as large and small components respectively.

\section{\label{cc-overview}Overview of the coupled cluster theory : method
of calculation}

We start with an $N$ electron closed shell Dirac-Fock (DF) reference
state $\left|\Phi\right\rangle $. The corresponding correlated closed
shell state is then

\begin{equation}
\left|\Psi\right\rangle =\exp(T)\left|\Phi\right\rangle ,\label{cc-1}\end{equation}
where $T$ is the core electron excitation operator. Then the Dirac-Coulomb
eigenvalue equation is 

\begin{equation}
H\exp(T)\left|\Phi\right\rangle =E\exp(T)\left|\Phi\right\rangle ,\label{cc-2}\end{equation}
with the Dirac-Coulomb Hamiltonian

\begin{equation}
H=\sum_{i}\left(c\alpha_{i}\cdot p_{i}+(\beta_{i}-1)mc^{2}+V_{N}\right)+\sum_{i<j}\frac{1}{r_{ij}}.\label{cc-3}\end{equation}
This leads to the exact ground state energy $E$ of the closed-shell
part of the system. Here $\alpha_{i}$ and $\beta_{i}$ are Dirac
matrices and $V_{N}$ is the nuclear potential. If we consider the
DF state $\left|\Phi\right\rangle $ as the Fermi vacuum, then the
normal ordered Hamiltonian is 

\begin{equation}
H_{N}\equiv H-\left\langle \Phi\right|H\left|\Phi\right\rangle =H-E_{DF}.\label{cc-4}\end{equation}

If we project $\left\langle \Phi\right|\exp(-T)$ from the left we
obtain the correlation energy ($\Delta E$) and if we project any
of the excited determinant $\left\langle \Phi^{\star}\right|\exp(-T)$
we additionally get a set of equations which are used to obtain the
$T$ amplitudes. Using the normal ordered dressed Hamiltonian $\overline{H_{N}}=\exp(-T)H_{N}\exp(T)$
the corresponding equations for correlation energy and amplitudes
become

\begin{equation}
\left\langle \Phi\right|\overline{H_{N}}\left|\Phi\right\rangle =\Delta E,\label{cc-5}\end{equation}
and

\begin{equation}
\left\langle \Phi^{\star}\right|\overline{H_{N}}\left|\Phi\right\rangle =0.\label{cc-6}\end{equation}

Here the state $\left|\Phi^{\star}\right\rangle $ may be singly excited
$\left|\Phi_{a}^{r}\right\rangle $ or double excited $\left|\Phi_{ab}^{rs}\right\rangle $
and so on. The indices $a,b,\cdots$ refer to holes and $p,q,\cdots$
to particles. We have considered the coupled cluster single and double
(CCSD) approximation, where the cluster operator $T$ is composed
of one- and two-body excitation operators, \emph{i.e.} $T=T_{1}+T_{2}$,
and are expressed in second quantization form

\begin{equation}
T=T_{1}+T_{2}=\sum_{ap}a_{p}^{\dagger}a_{a}t_{a}^{p}+\frac{1}{2}\sum_{abpq}a_{p}^{\dagger}a_{q}^{\dagger}a_{b}a_{a}t_{ab}^{pq}.\label{cc-7}\end{equation}

Contracting the ladder operators \cite{bartlett-book} and rearranging
the indices, the amplitude equations can be expressed in the form

\begin{equation}
A+B(T)\cdot T=0,\label{cc-8}\end{equation}
where $A$ is a constant vector consisting of the matrix elements
$\left\langle \Phi^{\star}\right|H_{N}\left|\Phi\right\rangle $,
$T$ is the vector of the excitation amplitudes and $B(T)$ is the
matrix which depends on the cluster amplitudes itself so that Eq.
(\ref{cc-8}) is solved self-consistently. For example, a typical
contribution to the term $\widehat{\widehat{H_{N}T_{2}}T_{2}}$ is

\begin{equation}
B_{ab}^{pq}=\frac{1}{2}\sum_{dgrs}V_{dgrs}t_{ad}^{pr}t_{gb}^{sq}.\label{cc-9}\end{equation}
Here $V_{dgrs}$ is the two-electron Coulomb integral and $t_{ad}^{pr}$
is the cluster amplitude corresponding to a simultaneous excitation
of two electrons from orbital $a$ and $d$ to $p$ and $r$ respectively.
To obtain a full set of terms which contribute to this specific excitation,
diagrammatic techniques are used.

The ground state of $^{25}Mg^{+}$ contains only one valance electron
in the outer most orbital ($3s_{1/2}$). To calculate the ground state
energy of the system we first compute the correlations for the closed
shell system ($^{25}Mg^{+2}$) using the closed shell coupled cluster
approach and then use the technique of electron attachment (open shell
coupled cluster (OSCC)) method. The energy of the excited state are
obtained by the same way. In order to add an electron to the $k$th
virtual orbital of the DF reference state we define

\begin{equation}
\left|\Phi_{k}^{N+1}\right\rangle \equiv a_{k}^{\dagger}\left|\Phi\right\rangle \label{cc-10}\end{equation}
with the particle creation operator $a_{k}^{\dagger}$. Then by using
the excitation operators for both the core and valance electron the
exact state is defined as \cite{lindgren-book}:

\begin{equation}
\left|\Psi_{k}^{N+1}\right\rangle =\exp(T)\left\{ \exp(S_{k})\right\} \left|\Phi_{k}^{N+1}\right\rangle .\label{cc-11}\end{equation}
Here $\left\{ \exp(S_{k})\right\} $ is the normal ordered exponential
representing the valance part of the wave operator. Here 

\begin{equation}
S_{k}=S_{1k}+S_{2k}=\sum_{k\neq p}a_{p}^{\dagger}a_{k}s_{k}^{p}+\frac{1}{2}\sum_{bpq}a_{p}^{\dagger}a_{q}^{\dagger}a_{b}a_{k}s_{kb}^{pq}\,,\label{s-expr}\end{equation}
where $k$ stands for valance orbital. $S_{k}$ contain the particle
annihilation operator $a_{k}$, and because of the normal ordering
it cannot be connected to any other valance electron excitation operator
and then $\left\{ \exp(S_{k})\right\} $ automatically reduces to
$\left\{ 1+S_{k}\right\} $. 

Then we can write the Eq.(\ref{cc-11}) as 

\begin{equation}
\left|\Psi_{k}^{N+1}\right\rangle =\exp(T)\left\{ 1+S_{k}\right\} \left|\Phi_{k}^{N+1}\right\rangle .\label{cc-12}\end{equation}
Following the same procedure as in the closed-shell approach, we obtain
a set of equations 

\begin{equation}
\left\langle \Phi_{k}^{N+1}\right|\overline{H_{N}}\left\{ 1+S_{k}\right\} \left|\Phi_{k}^{N+1}\right\rangle =H_{eff}\label{cc-13}\end{equation}
and

\begin{equation}
\left\langle \Phi_{k}^{^{\star}N+1}\right|\overline{H_{N}}\left\{ 1+S_{k}\right\} \left|\Phi_{k}^{N+1}\right\rangle =H_{eff}\left\langle \Phi_{k}^{^{\star}N+1}\right|\left\{ 1+S_{k}\right\} \left|\Phi_{k}^{N+1}\right\rangle ,\label{cc-14}\end{equation}
where the desired roots can be obtained by diagonalizing $H_{eff}$.
The Eq.(\ref{cc-14}) is non-linear in $S_{k}$ because the energy
difference $H_{eff}$ is itself a function of $S_{k}$. Hence, these
equations have to solved self-consistently to determine the $S_{k}$
amplitudes. 

\begin{figure}
\begin{center}\includegraphics[%
  scale=0.5]{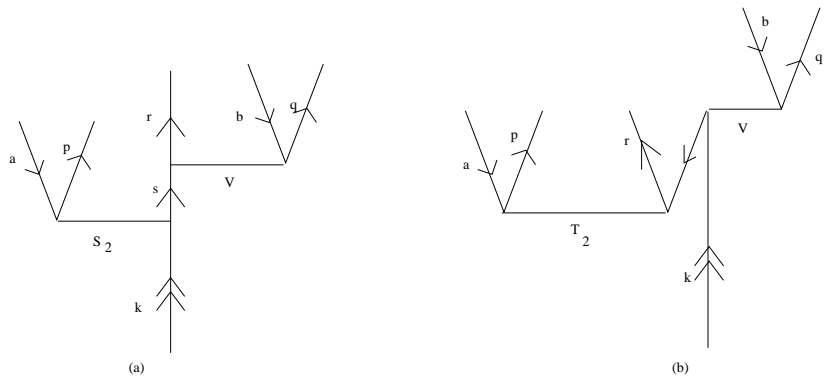}\end{center}

\caption{\label{diag-triples}Some typical important diagrams which arise
due to the inclusion of triples through Eq.(\ref{cc-15})}
\end{figure}

Triple excitations are included in our open shell CC amplitude calculations
by an approximation that is similar in spirit to CCSD(T) \cite{ccsd(t)}.
The approximate triple excitation amplitude is given by

\begin{equation}
S_{abk}^{pqr}=\frac{\widehat{VT_{2}}+\widehat{VS_{2}}}{\varepsilon_{a}+\varepsilon_{b}+\varepsilon_{k}-\varepsilon_{p}-\varepsilon_{q}-\varepsilon_{r}},\label{cc-15}\end{equation}
where $S_{abk}^{pqr}$ are the amplitudes corresponding to the simultaneous
excitation of orbitals $a,b,k$ to $p,q,r$ respectively and $\widehat{VT}$
and $\widehat{VS}$ are the correlated composites involving $V$ and
$T$, and $V$ and $S$ respectively. $\varepsilon_{k}$ is the orbital
energy of the $k$th orbital. The above amplitudes (some representative
diagrams are given in figure \ref{diag-triples}) are added appropriately
in the singles and doubles open shell cluster amplitude equations
and these equations are then solved self-consistently. We therefore
obtain solutions of $S_{1}$and $S_{2}$ amplitudes taking into consideration
the effect of the triple excitations in an approximate way

The expectation value of any operator $O$ can be written as the normalized
form with respect to the exact state $\left|\Psi^{N+1}\right\rangle $
as

\begin{equation}
\left\langle O\right\rangle =\frac{\left\langle \Psi^{N+1}\right|O\left|\Psi^{N+1}\right\rangle }{\left\langle \Psi^{N+1}\right|\left.\Psi^{N+1}\right\rangle }=\frac{\left\langle \Phi^{N+1}\right|\left\{ 1+S^{\dagger}\right\} \exp(T^{\dagger})O\exp(T)\left\{ 1+S\right\} \left|\Phi^{N+1}\right\rangle }{\left\langle \Phi^{N+1}\right|\left\{ 1+S^{\dagger}\right\} \exp(T^{\dagger})\exp(T)\left\{ 1+S\right\} \left|\Phi^{N+1}\right\rangle }.\label{cc-16}\end{equation}

For computational simplicity we store only the one-body matrix element
of $\overline{O}=\exp(T^{\dagger})O\exp(T)$. $\overline{O}$ may
be expressed in terms of uncontracted single-particle lines \cite{geetha}.
The fully contracted part of $\overline{O}$ will not contribute as
it cannot be linked with the remaining part of the numerator of the
above equation.

In the LCCSD approximation Eq. (\ref{cc-11}) turns out to be 

\begin{equation}
\left|\Psi_{k}^{N+1}\right\rangle =\left\{ 1+T+S_{k}\right\} \left|\Phi_{k}^{N+1}\right\rangle ,\label{lccsd-open}\end{equation}
and $\overline{H_{N}}=H_{N}+\overbrace{H_{N}T}$. 

The closed and open shell cluster amplitude equations reduce to

\begin{equation}
\left\langle \Phi_{0}^{k}\right|H_{N}+\overbrace{H_{N}T}\left|\Phi_{0}\right\rangle =0,\label{lccsd-closed}\end{equation}
and

\begin{equation}
\left\langle \Phi_{k}^{^{\star}N+1}\right|\overline{H_{N}}\left\{ 1+S_{k}\right\} \left|\Phi_{k}^{N+1}\right\rangle =\left\langle \Phi_{k}^{^{\star}N+1}\right|S_{v}\left|\Phi_{k}^{N+1}\right\rangle \left\langle \Phi_{k}^{N+1}\right|H_{N}\left|\Phi_{k}^{N+1}\right\rangle .\label{lccsd-1}\end{equation}

The orbitals used in the present work are expanded in terms of a finite
basis set comprising of Gaussian type orbitals (GTO) \cite{napp-fbse}

\begin{equation}
F_{i,k}(r)=r^{k}\exp(-\alpha_{i}r^{2}),\label{comp-1}\end{equation}
with $k=0,1,2\cdots$ for $s,p,d,\cdots$ type functions, respectively.
The exponents are determined by the even tempering condition \cite{even-tem}

\begin{equation}
\alpha_{i}=\alpha_{0}\beta^{i-1}.\label{comp-2}\end{equation}

The staring point of the computation is the generation of the Dirac-Fock
(DF) orbitals \cite{napp-fbse} which are defined on a radial grid
of the form

\begin{equation}
r_{i}=r_{0}\left[\exp(i-1)h-1\right]\label{comp-3}\end{equation}
with the freedom of choosing the parameters $r_{0}$ and $h$. All
DF orbitals are generated using a two parameter Fermi nuclear distribution

\begin{equation}
\rho=\frac{\rho_{0}}{1+\exp((r-c)/a)}\,,\label{fermi-nucl}\end{equation}
 where the parameter $c$ is the half charge radius and $a$ is related
to skin thickness, defined as the interval of the nuclear thickness
in which the nuclear charge density falls from near one to near zero.

Although we have used a large basis for the generation of the single
particle orbitals, the high-lying virtual orbitals (above a certain
threshold) are kept frozen as their contributions to the high-lying
virtuals in the $T$ and $S$ amplitudes in the CC equations are negligible.
Another advantage of this approximation is that it reduces the memory
required to store the matrix elements of the dressed operator $\overline{H}$
and the two-electron Coulomb integrals in the main memory, thereby
reducing the computational cost. In our calculations, we have included
all possible single, double and partial triple excitations from the
core.

We have used different $\alpha_{0}$ and $\beta$ for different symmetries.
The number of basis functions used to generate the even tempered DF
states are listed in table \ref{gauss-table} and the values of the
parameters $\alpha_{0}$ and $\beta$ used are also listed. For coupled
cluster calculations, we have restricted the basis by imposing an
upper bound in energy for single particle orbitals and the convergence
of our results are shown in table \ref{basis-conv}.

\begin{figure}
\begin{center}\includegraphics{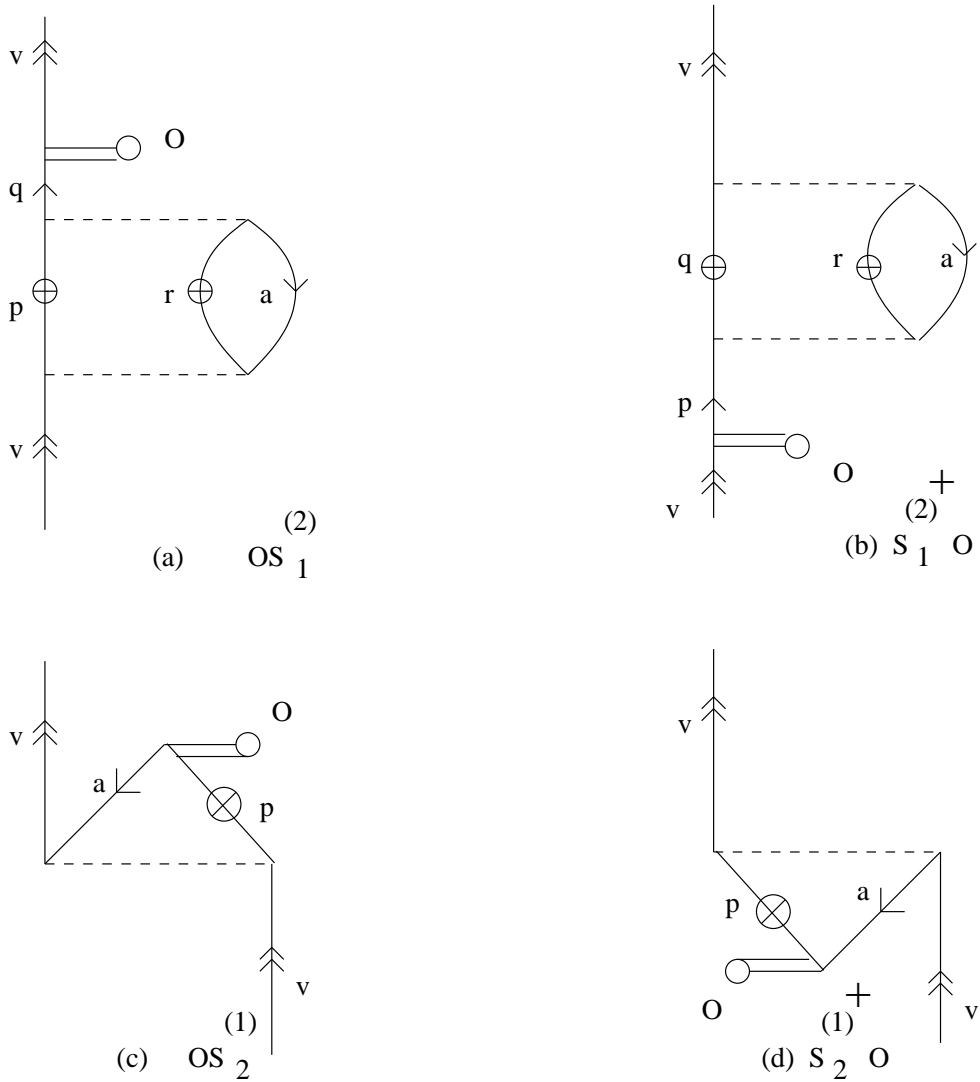}\end{center}

\caption{\label{diag-mbpt}Some typical important MBPT diagrams for pair correlation
and core-polarization effects. The superscripts refer to the order
of perturbation and the dashed lines correspond to the Coulomb interaction.
Particles and holes (labeled by $a$) are denoted by the lines directed
upward and downward respectively. The double line represents the O
(the hyperfine interaction operator) vertices. The valance (labeled
by $v$) and virtual orbitals (labeled by $p,q,r..$) are depicted
by double arrow and single arrow respectively, whereas the orbitals
denoted by $\oplus$ can either be valance or virtual.}
\end{figure}
\begin{figure}
\begin{center}\includegraphics{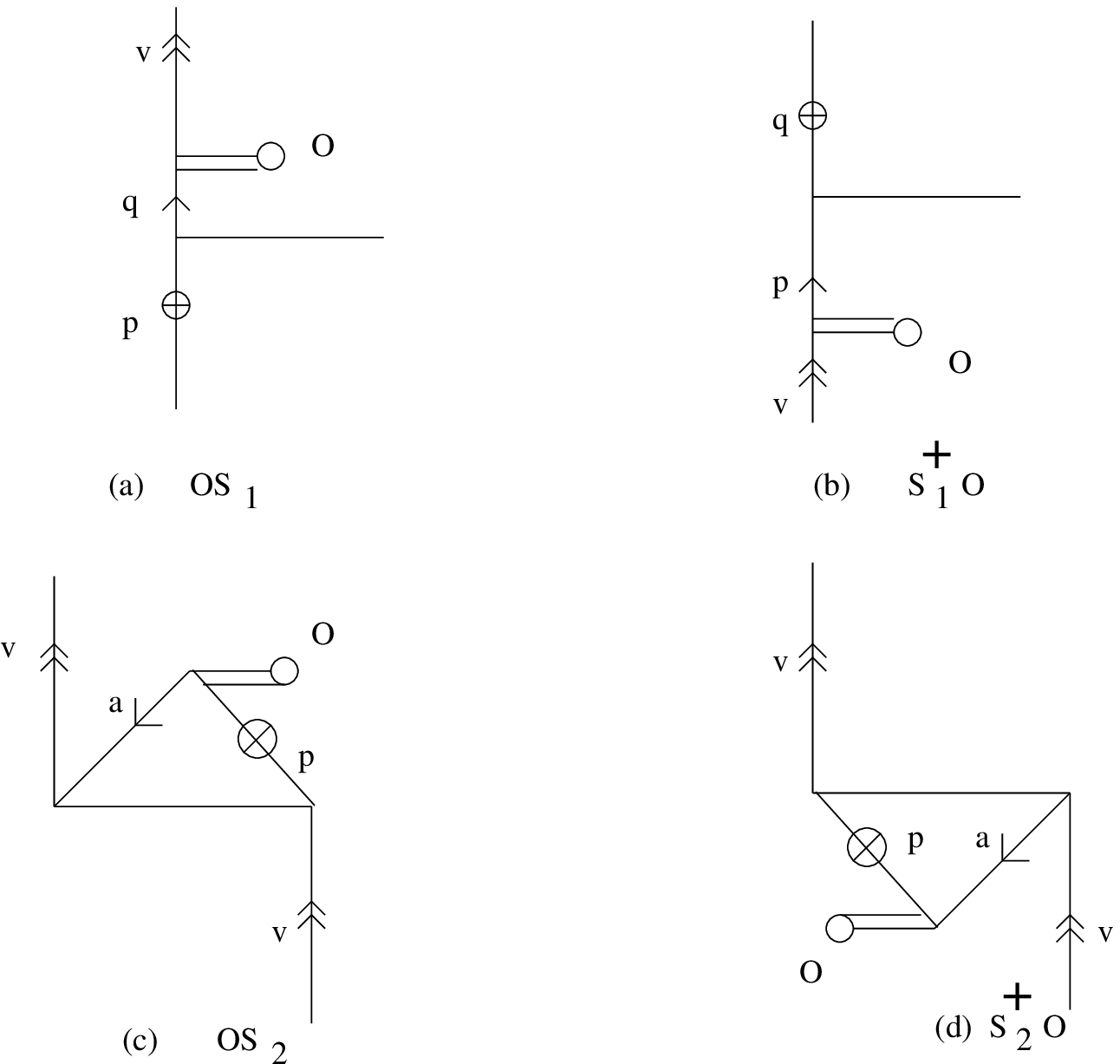}\end{center}

\caption{\label{diag-cc}The corresponding CC diagrams for pair correlation
and core-polarization effects. Here $a$ denotes a hole whereas $v$
denotes valance orbital and $p,q,r...$ denote virtual orbitals (particles).}
\end{figure}

\section{\label{results}Results and discussions}

The values of the magnetic dipole hyperfine constant $A$ and electric
quadrupole hyperfine constant $B$ for different states are given
in table \ref{Mg+-hyp-A-B}. Our calculated value of $A$ for the
ground state $3s_{1/2}$ is in good agreement (less than 0.6\%) with
experiment and it is more accurate than a previous calculation ($\sim$1\%)
based on second order relativistic many-body perturbation theory (RMBPT)
\cite{tpdas-mg+}. This is because unlike the previous work our calculation
is based on an approach which is equivalent to all-order MBPT. In
particular, we have taken into account all single, double and a subset
of triple excitations to all orders in the residual Coulomb interaction.
It is, therefore, not surprising that the result of our calculation
when carried out by using second order RMBPT is in agreement with
the result of the earlier calculation (see table \ref{Mg+-hyp-A-B})
\cite{tpdas-mg+}.

In table \ref{Mg+-hyp-A-B} the results calculated by Safronova \emph{et
al} \cite{johnson-hyp} using linearised coupled cluster in singles
doubles approximation (LCCSD) are given. We have performed LCCSD(T)
calculations and the results are given in the same table. The reasons
for the discrepancy between the two linearised coupled-cluster calculations
seem to be the inclusion of partial triple excitations by us and the
different choice of basis sets. From our present calculations it is
possible to determine the contributions from the non-linear clusters.
These contributions vary from 0.3\% ($3s_{1/2}$) to 0.8\% ($3d_{3/2}$)
for the different states. We find that if we take the linear contributions
of the $T$ amplitudes but the nonlinear contributions of the combined
$T$ and $S$ amplitudes and then perform the calculation of the hyperfine
constants (Eq.\ref{cc-16}) at the linear level (which we have named
as (L)CCSD(T)), the result ($A(3s_{1/2})=597.45$ MHz) is in excellent
agreement with the experiment (error is $\sim$0.2\%) for the ground
state. But theoretically this approach is not complete and hence a
proper inclusion of the nonlinear effects is desirable. Table \ref{ccsdVSccsdt}
shows the contribution of triples in the final property. Some partial
triples effect are taken into account iteratively according to Eq.
(\ref{cc-15}) and the $S$ amplitude thus contain some partial triples.
The effect of triples in calculation of properties (Eq. \ref{cc-16})
thus comes from the triples effect included in $S$ amplitude. Figure
\ref{diag-triples} shows some of the typical diagrams which give
rise to the effect due to triples, where figure \ref{diag-triples}a
and \ref{diag-triples}b correspond to the terms $\widehat{VS_{2}}$
and $\widehat{VT_{2}}$ respectively.

\begin{table}

\caption{\label{gauss-table}No. of basis functions used to generate the even
tempered Dirac-Fock orbitals and the corresponding value of $\alpha_{0}$
and $\beta$ used.}

~

\begin{center}\begin{tabular}{cccccccc}
\hline 
&
$s_{1/2}$&
$p_{1/2}$&
$p_{3/2}$&
$d_{3/2}$&
$d_{5/2}$&
$f_{5/2}$&
$f_{7/2}$\tabularnewline
\hline
\hline 
Number of basis&
35&
32&
32&
25&
25&
25&
25\tabularnewline
$\alpha_{0}$&
0.00625&
0.00638&
0.00638&
0.00654&
0.00654&
0.00667&
0.00667\tabularnewline
$\beta$&
2.03&
2.07&
2.07&
2.19&
2.19&
2.27&
2.27\tabularnewline
\hline
\end{tabular}\end{center}
\end{table}

\begin{table}

\caption{\label{Mg+-hyp-A-B}Value of magnetic dipole (\emph{A}) and electric
quadrupole (\emph{B}) hyperfine constants in MHz for $^{25}Mg^{+}$.
(T) stands for CC with perturbative partial triples.}

~

\begin{center}\begin{tabular}{cllrrrrr}
\hline 
States &
CCSD(T)$^{(a)}$&
&
MBPT&
LCCSD(T)&
(L)CCSD(T)&
Others$^{(d)}$&
Experiment$^{(e)}$\tabularnewline
\hline
\hline 
&
A&
B&
A&
A$^{(c)}$&
A$^{(c)}$&
A&
A\tabularnewline
$3s_{1/2}$&
592.86&
&
602(8)$^{(b)}$ &
590.73&
597.45&
597.6&
596.2544(5)\tabularnewline
&
&
&
602.46$^{(c)}$&
&
&
&
\tabularnewline
$4s_{1/2}$&
162.32&
&
164.65$^{(c)}$&
161.79&
163.34&
163.4&
\tabularnewline
$3p_{1/2}$&
101.70&
&
103.20$^{(c)}$&
100.69&
102.40&
103.4&
\tabularnewline
$4p_{1/2}$&
33.83&
&
34.26$^{(c)}$&
33.51&
34.05&
&
\tabularnewline
$3p_{3/2}$&
18.89&
22.91&
19.94$^{(c)}$&
18.95&
19.02&
19.29&
\tabularnewline
$4p_{3/2}$&
6.21&
7.48&
6.54$^{(c)}$&
6.24&
6.25&
&
\tabularnewline
$3d_{3/2}$&
1.17&
1.26&
1.16$^{(c)}$&
1.16&
1.17&
1.140&
\tabularnewline
$4d_{3/2}$&
0.51&
0.48&
0.505$^{(c)}$&
0.51&
0.51&
&
\tabularnewline
\hline
\hline 
&
&
&
&
&
&
&
\tabularnewline
\end{tabular}\end{center}

Refs. : (\emph{a}) : Present work, (\emph{b}) \cite{tpdas-mg+}, (c)
Present work, (\emph{d}) : \cite{johnson-hyp}, (\emph{e}) : \cite{mg+exp}
\end{table}

\begin{table}

\caption{\label{basis-conv}Convergence of results for A ($3s_{1/2}$ state)
using different basis sets}

\begin{center}\begin{tabular}{ccccccccc}
\hline 
&
$s_{1/2}$&
$p_{1/2}$&
$p_{3/2}$&
$d_{3/2}$&
$d_{5/2}$&
$f_{5/2}$&
$f_{7/2}$&
$A$(in MHz)\tabularnewline
\hline
\hline 
Number &
11&
9&
9&
8&
8&
7&
7&
562.10\tabularnewline
of basis&
14 &
12 &
12&
10 &
10&
8&
8&
589.18\tabularnewline
&
16&
14&
14&
11&
11&
10&
10&
592.86\tabularnewline
&
17&
15&
15&
11&
11&
10&
10&
592.86\tabularnewline
\hline
\hline 
&
&
&
&
&
&
&
&
\tabularnewline
\end{tabular}\end{center}
\end{table}
\begin{table}

\caption{\label{ccsdVSccsdt}Comparative tables for the value of magnetic
dipole (\emph{A}) hyperfine constants in MHz for $^{25}Mg^{+}$ calculated
using CCSD and CCSD(T). The difference between the values demonstrate
the effect of triples.}

~

\begin{center}\begin{tabular}{cllr}
\hline 
States &
CCSD(T)&
CCSD&
Experiment\tabularnewline
\hline
$3s_{1/2}$&
592.86&
593.01&
596.2544(5)\tabularnewline
$4s_{1/2}$&
162.32&
162.39&
\tabularnewline
\hline
\hline 
&
&
&
\tabularnewline
\end{tabular}\end{center}
\end{table}

\begin{table}

\caption{\label{comp-table-1}Comparative study of the contribution from different
terms (CCSD(T)) containing the dressed operator $\overline{O}$ in
determining the value of magnetic dipole hyperfine constant \emph{A}
for $^{25}Mg^{+}$for the different states.}

~

\begin{center}\begin{tabular}{ccccccccc}
\hline 
Terms&
$3s_{1/2}$&
$4s_{1/2}$&
$3p_{1/2}$&
$4p_{1/2}$&
$3p_{3/2}$&
$4p_{3/2}$&
$3d_{3/2}$&
$4d_{3/2}$\tabularnewline
\hline
\hline 
$\overline{O}$&
468.819&
131.616&
77.975&
26.400&
15.337&
5.196&
1.262&
0.563\tabularnewline
$\overline{O}S_{1}+S_{1}^{\dagger}\overline{O}$&
40.046&
8.256&
7.344&
2.096&
1.442&
0.412&
0.070&
0.031\tabularnewline
$\overline{O}S_{2}+S_{2}^{\dagger}\overline{O}$&
77.002&
20.587&
14.891&
4.872&
1.832&
0.519&
-0.175&
-0.086\tabularnewline
$S_{1}^{\dagger}\overline{O}S_{1}$&
0.855&
0.129&
0.179&
0.044&
0.035&
0.009&
0.002&
0.0007\tabularnewline
$S_{2}^{\dagger}\overline{O}S_{1}$&
1.175&
0.162&
0.238&
0.048&
0.0004&
-0.007&
-0.003&
-0.001\tabularnewline
$S_{1}^{\dagger}\overline{O}S_{2}$&
1.175&
0.162&
0.238&
0.048&
0.0004&
-0.007&
-0.003&
-0.001\tabularnewline
$S_{2}^{\dagger}\overline{O}S_{2}$&
5.447&
1.613&
0.997&
0.350&
0.268&
0.101&
0.015&
0.008\tabularnewline
\hline
\end{tabular}\end{center}
\end{table}
\begin{table}

\caption{\label{comp-table-2}Comparative study of the most contributing terms
containing the operator $O$ from (CCSD(T)) in determining the value
of magnetic dipole hyperfine constant \emph{A} for $^{25}Mg^{+}$for
the different states.}

~

\begin{center}\begin{tabular}{ccccccccc}
\hline 
Terms&
$3s_{1/2}$&
$4s_{1/2}$&
$3p_{1/2}$&
$4p_{1/2}$&
$3p_{3/2}$&
$4p_{3/2}$&
$3d_{3/2}$&
$4d_{3/2}$\tabularnewline
\hline
\hline 
$O$&
468.819&
130.616&
77.975&
26.400&
15.337&
5.196&
1.262&
0.563\tabularnewline
$OS_{1}+S_{1}^{\dagger}O$&
39.713&
8.233&
7.293&
2.093&
1.440&
0.414&
0.070&
0.031\tabularnewline
$OS_{2}+S_{2}^{\dagger}O$&
77.767&
20.836&
15.153&
4.973&
1.984&
0.569&
-0.194&
-0.096\tabularnewline
\hline
\end{tabular}\end{center}
\end{table}

\begin{table}

\caption{\label{ind-cont}Individual contribution from the $OS_{1}$and $OS_{2}$
diagrams for $3s_{1/2}$ state. The values given correspond to the
respective terms in MHz.}

\begin{center}\begin{tabular}{ccccc}
\hline 
Orbital&
$\overline{O}S_{1}$&
$OS_{1}$&
$\overline{O}S_{2}$&
$OS_{2}$\tabularnewline
\hline
\hline 
$3s_{1/2}$&
20.023&
19.857&
&
\tabularnewline
&
&
&
&
\tabularnewline
$1s_{1/2}$&
&
&
13.309&
13.196\tabularnewline
$2s_{1/2}$&
&
&
26.665&
25.699\tabularnewline
$2p_{1/2}$&
&
&
0.882&
0.867\tabularnewline
$2p_{3/2}$&
&
&
-0.958&
-0.878\tabularnewline
\hline
\hline 
&
&
&
&
\tabularnewline
\end{tabular}\end{center}
\end{table}

For the present calculations, the number of basis functions actually
used are the following : 16$s_{1/2}$, 14$p$ ($p_{1/2}$and $p_{3/2}$),
11$d$ ($d_{3/2}$and $d_{5/2}$) and 10$f$ ($f_{5/2}$and $f_{7/2}$).
Excitations from all the core orbitals have been considered.

The important contributions to the magnetic dipole hyperfine constants
for different states are given in table \ref{comp-table-1}. In particular,
we have analyzed the contributions from various many body effects
and have demonstrated that the most important contributions come from
core polarization and pair correlation effects. The largest contribution
comes from $\overline{O}$. The next two largest contributions come
from $(\overline{O}S_{1}+S_{1}^{\dagger}\overline{O})$ and$(\overline{O}S_{2}+S_{2}^{\dagger}\overline{O})$
which correspond to the pair-correlation (PC) and core-polarization
(CP) effects respectively. The contribution from the corresponding
MBPT terms are listed in table \ref{comp-table-2}. Figures \ref{diag-mbpt}
and \ref{diag-cc} represents the pair-correlation and core-polarization
diagrams in MBPT and CC respectively. 

We have listed the contributions from the different terms containing
the dressed operator $\overline{O}$ in table \ref{comp-table-1}
and table \ref{comp-table-2} gives the contributions from the terms
containing the operator $O$ directly. The results given in tables
\ref{comp-table-1} and \ref{comp-table-2} show that the CP contribution
is larger than the PC in magnitude for all the states, although the
ratio of the two effects is not uniform. It is important to note that
the former contribution includes the hyperfine interaction of all
the core orbitals while only a specific valence orbital is involved
in this interaction for the latter (see Figure \ref{diag-cc}). These
individual contributions are presented in table \ref{ind-cont} for
the ground state of $Mg^{+}$. However, the ground state hyperfine
constant $A$ for $Ba^{+}$exhibits exactly the opposite behaviour
\cite{napp-hyp-ba+}. Even though $Ba^{+}$ has more core electrons
than $Mg^{+}$, the relativistic enhancement of the valence ($6s$)
magnetic dipole hyperfine interaction results in the value of PC exceeding
that of CP.

\section{Conclusion}

In this paper we have carried out \emph{ab-initio} relativistic coupled
cluster calculations of magnetic dipole ($A$) and electric quadrupole
($B$) hyperfine constants for the ground and some excited states
of $^{25}Mg^{+}$. Some contributions from partial triples are also
taken into account in our calculation. We have shown that the dominant
many-body contributions to these properties come from core-polarization
and pair-correlation effects.

In addition to comparing with the available experimental data we have
also predicted the values of $A$ and $B$ for a few states which
could be of interest in the future. Using ion-trapping and other experimental
techniques, it may be possible to measure both the magnetic dipole
and electric quadrupole hyperfine constants for different states of
$^{25}Mg^{+}$, thereby checking the accuracy of our calculations.
This would constitute an useful test of the validity of the coupled-cluster
theory in capturing the many-body effects in hyperfine interactions
in light atomic systems with a single valence electron.

\begin{verse}
\textbf{Acknowledgments} : This work was supported by the BRNS for
project no. 2002/37/12/BRNS. The computation were carried out on our
group's 4 CPU E450 Sun Ultra SPARC machine in IIA and on CDAC, Bangalore's
Teraflop Supercomputer Param Padma. 
\end{verse}

\end{document}